\documentclass{aa}  

\usepackage{color}

\usepackage{graphicx}

\usepackage{mathtools}
\usepackage{siunitx}
\usepackage[bottom]{footmisc}
\usepackage{tablefootnote}
\usepackage[flushleft]{threeparttable}
\usepackage{adjustbox}
\usepackage{subcaption}
\usepackage{float}

\usepackage{txfonts}
\begin{document}

   \title{MUSE observations of dwarf galaxies and a stellar stream in the M\,83 group\thanks{Based on observations collected at the European Organisation for Astronomical Research in the Southern Hemisphere under ESO program 112.262H.}}
   \author{Oliver M\"uller\inst{1,2,3}                             
   \and
         Marina Rejkuba\inst{4}
        \and         
        Katja Fahrion\inst{5,6}
        \and
         Marcel S. Pawlowski\inst{7}
         \and
        Benoit Famaey\inst{8}
         \and
        Noam Libeskind\inst{7}
        \and
         Nick Heesters\inst{1}
         \and
        Federico Lelli\inst{9}
         \and
         Michael Hilker\inst{4}
         \and
         Salvatore Taibi\inst{1,7}
         \and
         Sarah Pearson\inst{10}
          }
    
 \titlerunning{M83 dwarf galaxies observed with MUSE}
   \institute{
   Institute of Physics, Laboratory of Astrophysics, Ecole Polytechnique Fédérale de Lausanne (EPFL), 1290 Sauverny, Switzerland\\              
   \email{oliver.muller@epfl.ch}
   \and
   Institute of Astronomy, Madingley Rd, Cambridge CB3 0HA, UK
\and
Visiting Fellow, Clare Hall, University of Cambridge, Cambridge, UK
\and
 European Southern Observatory, Karl-Schwarzschild Strasse 2, 85748, Garching, Germany
 \and
 Department of Astrophysics, University of Vienna, T\"{u}rkenschanzstra{\ss}e 17, 1180 Wien, Austria
 \and
European Space Agency, European Space Research and Technology Centre, Keplerlaan 1, 2201 AZ Noordwijk, The Netherlands
\and
Leibniz-Institut f\"ur Astrophysik Potsdam (AIP), An der Sternwarte 16, D-14482 Potsdam, Germany
\and
Observatoire Astronomique de Strasbourg  (ObAS), Universite de Strasbourg - CNRS, UMR 7550 Strasbourg, France
\and
INAF, Arcetri Astrophysical Observatory, Largo E. Fermi 5, 50125 Florence, Italy
\and
DARK, Niels Bohr Institute, University of Copenhagen, Blegdamsvej 17, 2100 Copenhagen, Denmark
}
   \date{Received tba; accepted tba}

 
  \abstract
   {{Spectroscopy for faint dwarf galaxies outside of our own Local Group is challenging. Here, we present MUSE spectroscopy to study the properties of four known dwarf satellites and one stellar stream (KK\,208) surrounding the nearby grand spiral M\,83, which resides together with the lenticular galaxy Cen\,A in the Centaurus group. This data complete the phase-space information for all known dwarf galaxies around M\,83 down to a completeness of $-$10 mag in the $V$ band. All studied objects have an {intermediate to} old and metal-poor stellar population and follow the stellar luminosity-metallicity relation as defined by the Local Group dwarfs. For the stellar stream  we serendipitously identify a previously unknown globular cluster, which is old and metal-poor. Two dwarf galaxies {(NGC\,5264 and dw1341-29)} may be a bound satellite of a satellite system due to their proximity and shared velocities. Having access to the positions and velocities of 13 dwarfs around M\,83, we estimate the mass of the group with different estimators. Ranging between 1.3 and $3.0 \times 10^{12}$\,M$_\odot$ for the halo mass we find it to be larger than previously assumed. This may impact the previously reported tension {for cold dark matter cosmology} with the count of dwarf galaxies. In contrast to Cen\,A, we do not find a co-rotating plane-of-satellites around M\,83. 
   }
   }

   \keywords{Galaxies: dwarf; galaxies: groups: individual: M83; galaxies: distances and redshifts; cosmology: large-scale structure of Universe.
               }
   \maketitle
%

\section{Introduction}

{The Centaurus group is one of the closest neighbors from our own Local Group. It is made up of two giant galaxies, the lenticular galaxy Cen\,A at 3.8\,Mpc \citep{2010PASA...27..457H} and the grand design spiral galaxy M\,83 at 4.9\,Mpc \citep{2009AJ....138..332J}. Both come with their own set of dwarf galaxy satellites. Past surveys doubled the number of known dwarf galaxies around Cen\,A \citep{2014ApJ...795L..35C,2016ApJ...823...19C,2017A&A...597A...7M,2018ApJ...867L..15T} and increased the numbers by 30\% around M\,83 \citep{2015A&A...583A..79M}, while putting the completeness of known dwarfs on par with the classical regime of dwarf galaxies within our Local Group (down to $M_V\approx-10$ mag). The confirmation of  dwarf galaxy membership was mainly done by measuring the tip of the red giant branch \citep{MuellerTRGB2018,MuellerTRGB2019,2019ApJ...872...80C}, getting distances with an accuracy  of 5 to 10\%. Other characterizations of properties, such as their velocities or stellar populations, require spectroscopy, which is a much more time-consuming task, especially for the quenched dwarf spheroidals without emission lines. Fortunately, with facilities such as the Very Large Telescope (VLT) it is within reach to get the needed spectroscopy to measure these properties.}

{The Cen\,A galaxy group, being the closer and more populated galaxy group than  the M\,83 group, has been used for extending near-field cosmology tests. \citet{MuellerTRGB2019} compared the abundance of the dwarf galaxies around Cen\,A with expectations from cosmological simulations and found an agreement within 2$\sigma$. More puzzling though in the context of $\Lambda$ cold dark matter ($\Lambda$CDM) cosmology is the arrangement and motion of dwarf galaxies around Cen\,A: The dwarf galaxies are arranged in a thick disk \citep{2015ApJ...802L..25T,Muller2016}, in which the dwarfs are co-moving based on their line-of-sight velocities \citep{2018Sci...359..534M,Muller2021b}, which may indicate co-rotation \citep{2023MNRAS.519.6184K}. This is at odds with $\Lambda$CDM at the 3$\sigma$ level. This is reminiscent to the Local Group where two such co-rotating planes have been discovered around the Milky Way \citep{2005A&A...431..517K,2008ApJ...680..287M,2012MNRAS.423.1109P} and the Andromeda galaxy \citep{2006AJ....131.1405K,2006MNRAS.365..902M,2013Natur.493...62I} and pose a challenge to $\Lambda$CDM (see e.g. the reviews by \citealt{2021NatAs...5.1185P} and \citealt{2022NatAs...6..897S}).}

{M\,83 is farther and less populated than Cen\,A, making such cosmological tests more difficult. Nevertheless, \citet{2024A&A...684L...6M} compared the abundance of the dwarf galaxies around M\,83 with expectations from $\Lambda$CDM. In this case, however, the count was too high compared to the expectations at a 3 to 5$\sigma$ level, depending how the $\Lambda$CDM predictions are made (either through comparison to cosmological simulations or using the stellar-halo mass relation). 

\citet{MuellerTRGB2018} suggested the existence of a plane-of-satellites around M\,83 based on the 3D distribution of dwarf galaxies, but noted that the uncertainties are as large as the dimensions of this plane. However, this putative plane is seen edge-on, meaning that if it is indeed a rotationally supported planar structure, we would find a signature with line-of-sight velocity observations, namely that the dwarfs on one side should be redshifted, and on the other side blueshifted, with respect to the groups mean motion. Such velocity information was only available for 8 dwarf galaxies and was inconclusive. Here we present follow up observations with the VLT to acquire accurate spectroscopy and derive the line-of-sight velocities of the five remaining known dwarf galaxies (including a tidally disrupting one). {Except dw1341-29, all of them were already confirmed members of the M\,83 based on distance measurements through the tip of the red giant branch.} {The spectroscopy also allows to estimate the stellar population properties of these targets and search for compact stellar objects such as star clusters.}}

{The manuscript is organized as follows: in Section\,\ref{sec:observations} we present the observations and data reduction, in Section\,\ref{sec:lum_metal} the luminosity-metallicity relation, in Section\,\ref{sec:glob} the  discovery of a globular cluster, in Section\,\ref{kinematics} the phase-space distribution of the dwarf galaxies, in Section\,\ref{sec:mass} several mass estimations for the M\,83 group, and finally in Section\,\ref{sec:disc} we discuss the results and draw our conclusions.}

\section{Observations, data reduction, and spectroscopy}
\label{sec:observations}

\begin{figure*}[ht]
    \centering
        \includegraphics[width=0.19\linewidth]{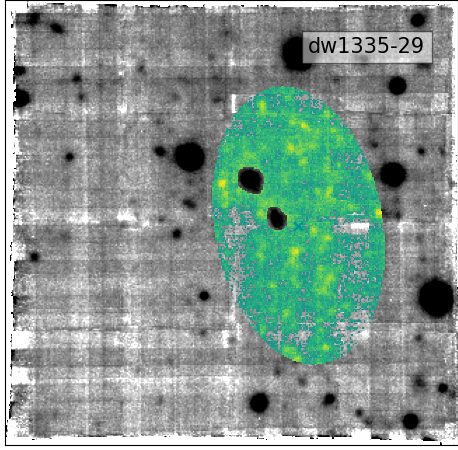}
        \includegraphics[width=0.19\linewidth]{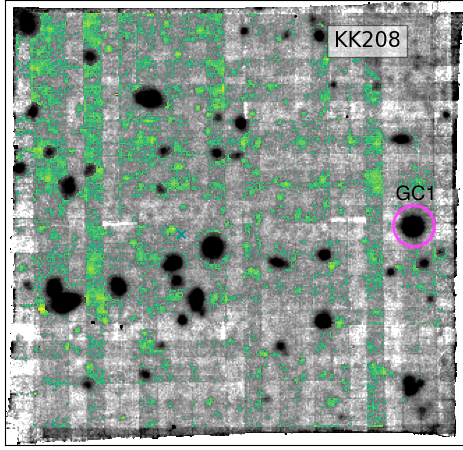}
        \includegraphics[width=0.19\linewidth]{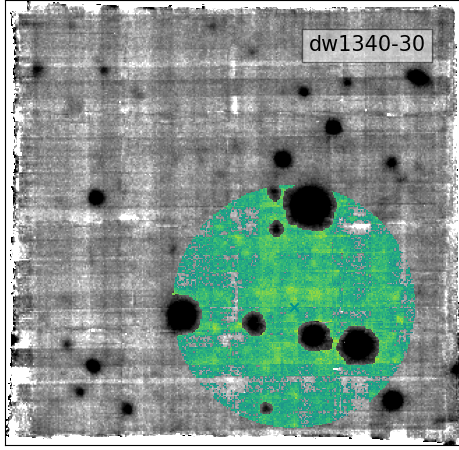}
        \includegraphics[width=0.19\linewidth]{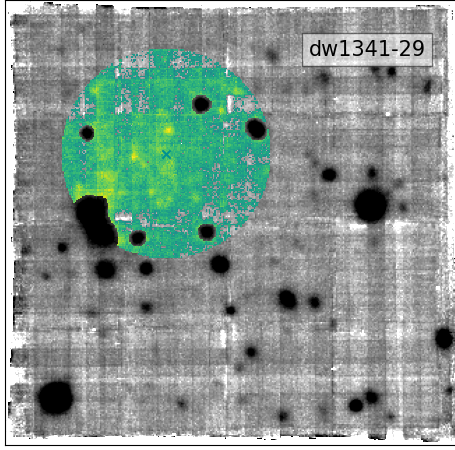}
        \includegraphics[width=0.19\linewidth]{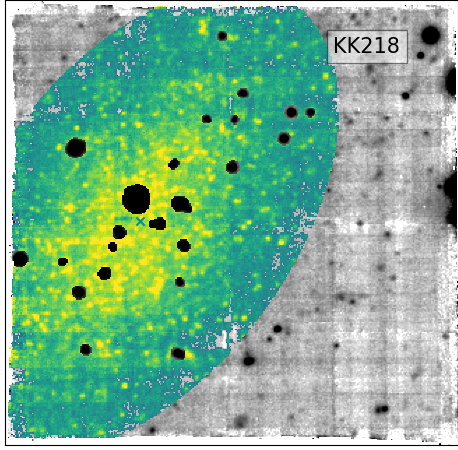}
    \caption{The stacked MUSE cubes. The colored areas indicate the regions where the spectra were extracted. The magenta circle indicates the GC discovered around KK\,208.
    }
    \label{fig:stacked}
\end{figure*}

The data were collected through the European Southern Observatory (ESO) program 112.262H (PI: Müller) using the Multi Unit Spectroscopic Explorer \citep[MUSE;][]{Bacon2010,2012Msngr.147....4B} on UT4 of the VLT at Cerro Paranal, Chile. MUSE is an integral field spectrograph (IFS) with a 1$\times$1 arcmin$^{2}$ field of view, spatial sampling of 0.2 arcsec/pixel, a nominal wavelength range of 480-930 nm, and a resolving power between 1740 (480 nm) and 3450 (930 nm). From a list of confirmed dwarf galaxies around M\,83 \citep{2024A&A...684L...6M}, we observed the remaining five dwarf galaxies without velocity information, listed in Table\,\ref{tab:sample}. 
Note that for one dwarf galaxy, dw1341-29, the value from surface brightness fluctuation measurements is closer to the distance of Cen\,A, but the large uncertainties makes it plausible that it is a member of M\,83 as well. Here, we assume it is a satellite member of M\,83, based on its projected proximity to M\,83 and its derived velocity being equal to a nearby dwarf (making it a potential satellite of satellite, see below).

For one object -- the tidally disrupted
dwarf KK\,208 -- we took additional sky exposures because the target filled the MUSE field-of-view. The total exposure time for this object was 1813s and 453s for the target and sky stacks, respectively. For the remaining objects, we observed on target for a total of 2559s with no extra sky exposures.
The MUSE data products, processed through the standard MUSE pipeline \citep{Weilbacher2012,2020arXiv200608638W}, are available via the ESO Science Archive. This processing included bias and flat-field corrections, astrometric calibration, sky subtraction, and wavelength and flux calibration \citep{Hanuschik2017}\footnote{see also \url{http://www.eso.org/observing/dfo/quality/PHOENIX/MUSE/processing.html}}.

For sky subtraction, we used the data cube directly where possible for the dwarf galaxies which are small enough to leave sufficient space for sky estimation, eliminating the need for separate sky exposures. \citet{Muller2021a} suggested that with 20\% of the MUSE field available for sky subtraction, better results are achieved using the science data cube instead of including a separate sky offset. {For KK\,208} we still got better results by deriving the sky on a small area within the science exposure rather than the sky exposure. 
To reduce the sky residual lines, we applied the Zurich Atmosphere Purge (ZAP) principal component analysis algorithm \citep{Soto2016}. We used the Python implementation \citep[][SEP]{SEP} of Source Extractor \citep{SExtractor} with a sigma threshold of 0.5 to select empty sky patches, creating a segmented FITS file of all detected sources. Additionally, we added a generous elliptical mask over the dwarf galaxies, because often SEP did not segment their outskirts. This segmentation map was used as a mask for ZAP.  

To extract the integrated spectrum of the galaxy body, we used an elliptical aperture adjusted on the collapsed cubes and used SEP to mask foreground stars and background objects, with threshold being between one to ten sigma. Additionally, we manually masked objects as needed. Further, we masked regions which had a median value of 0.1 or lower. This removes empty regions without any signal. The centroid, ellipticity and position angle of the ellipse were chosen by hand on the white image (i.e. the stacked data cube). The white image and the aperture are shown in Fig.\,\ref{fig:stacked}.

\begin{figure*}[ht]
    \centering
        \includegraphics[height=2.3cm]{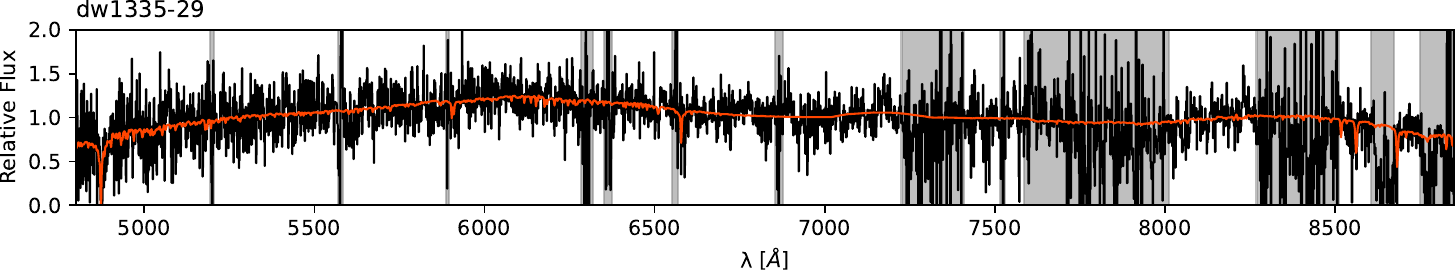}
        \includegraphics[height=2.3cm]{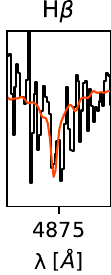}
        \includegraphics[height=2.3cm]{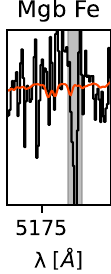}
        \includegraphics[height=2.3cm]{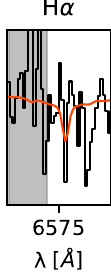}
        \includegraphics[height=2.3cm]{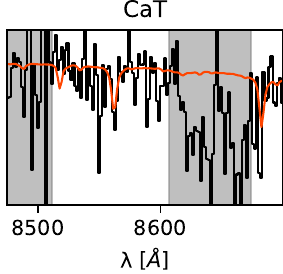}\\
        \includegraphics[height=2.3cm]{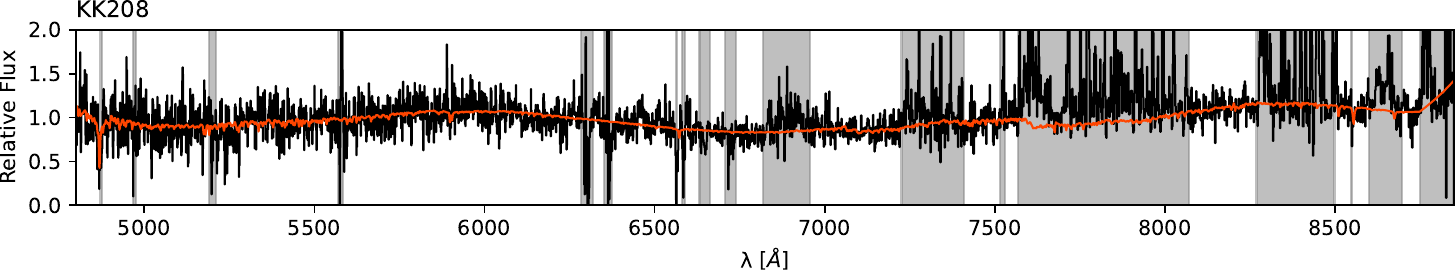}
        \includegraphics[height=2.3cm]{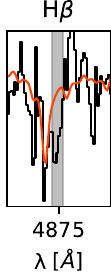}
        \includegraphics[height=2.3cm]{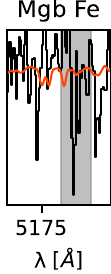}
        \includegraphics[height=2.3cm]{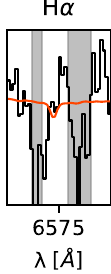}
        \includegraphics[height=2.3cm]{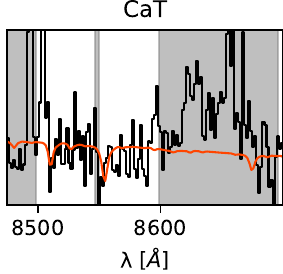}\\
        \includegraphics[height=2.3cm]{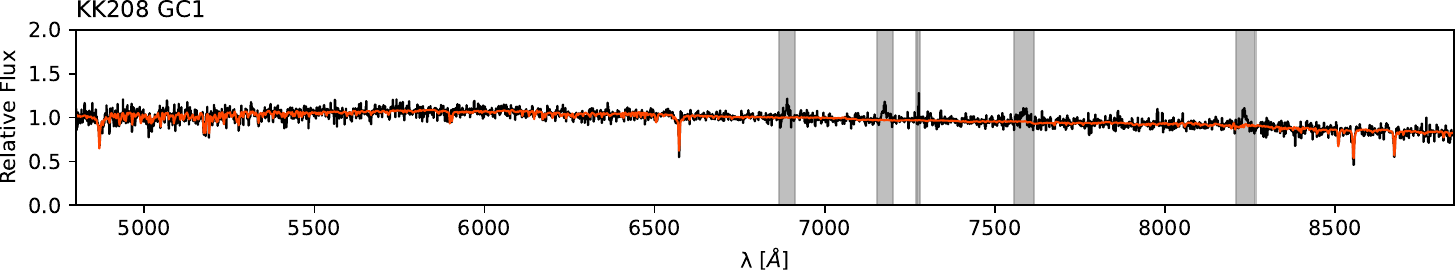}
        \includegraphics[height=2.3cm]{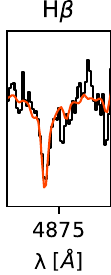}
        \includegraphics[height=2.3cm]{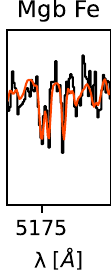}
        \includegraphics[height=2.3cm]{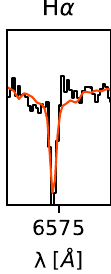}
        \includegraphics[height=2.3cm]{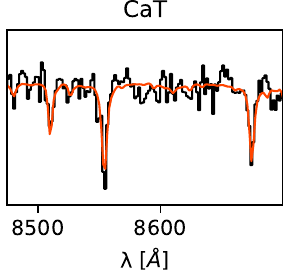}\\
        \includegraphics[height=2.3cm]{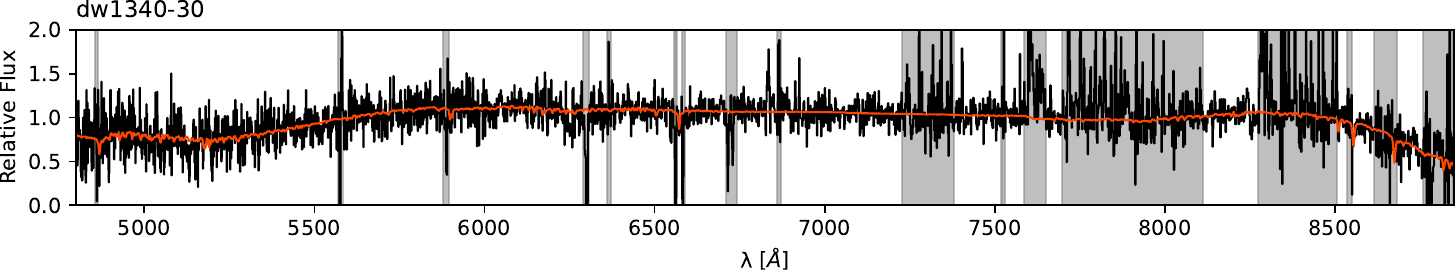}
        \includegraphics[height=2.3cm]{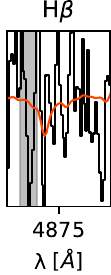}
        \includegraphics[height=2.3cm]{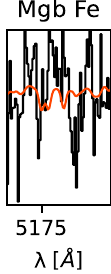}
        \includegraphics[height=2.3cm]{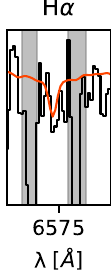}
        \includegraphics[height=2.3cm]{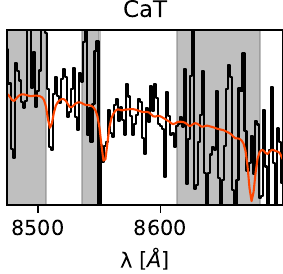}\\
        \includegraphics[height=2.3cm]{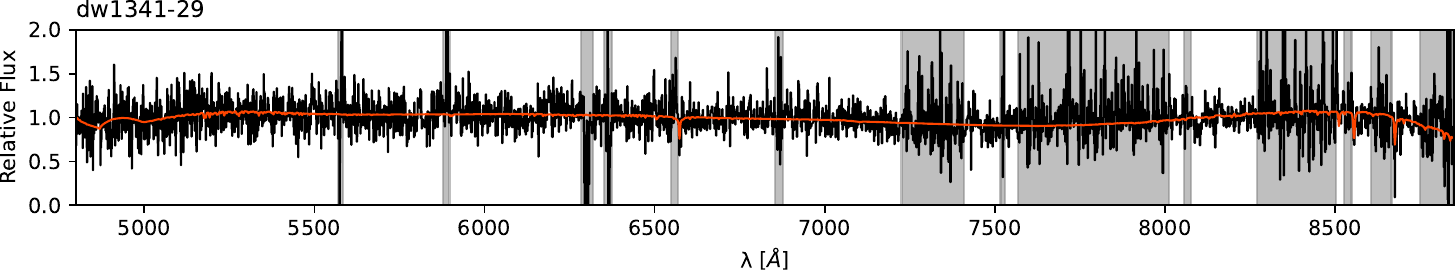}
        \includegraphics[height=2.3cm]{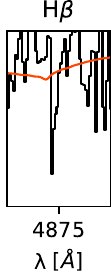}
        \includegraphics[height=2.3cm]{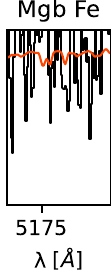}
        \includegraphics[height=2.3cm]{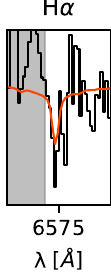}
        \includegraphics[height=2.3cm]{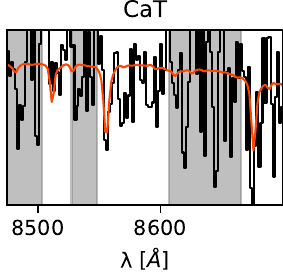}\\
        \includegraphics[height=2.3cm]{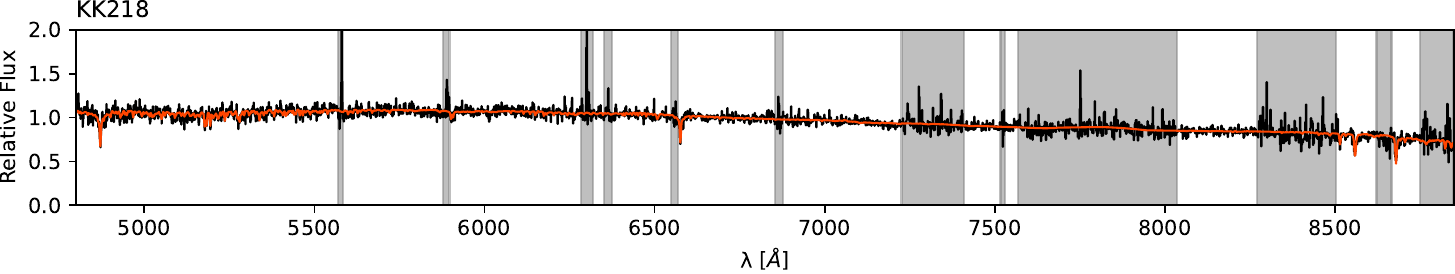}
        \includegraphics[height=2.3cm]{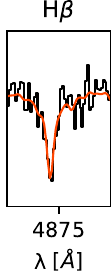}
        \includegraphics[height=2.3cm]{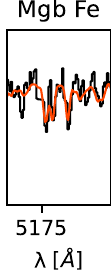}
        \includegraphics[height=2.3cm]{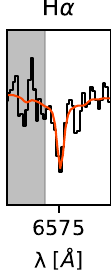}
        \includegraphics[height=2.3cm]{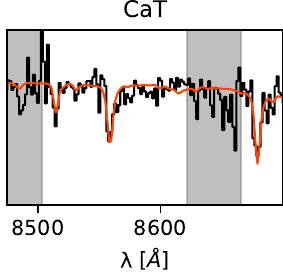}\\
    \caption{The spectra (black line) and the best fit from pPXF (red line) is plotted in the left panel over the full spectral coverage of MUSE. The gray area indicate masked regions. The middle and right panel show the two most prominent absorption line regions to derive the velocities, namely the region around $H\alpha$ and the Calcium Triplet (CaT). 
    }
    \label{fig:spectra}
\end{figure*}

To determine the line-of-sight velocities and stellar population properties, we employed the Python implementation of the penalized PiXel-Fitting (pPXF, \citealt{2004PASP..116..138C,2017MNRAS.466..798C}) algorithm, following the procedures used in previous studies with MUSE and pPXF {(see \citealt{2019A&A...625A..76E,2019A&A...625A..77F,2020A&A...640A.106M,2025A&A...693A..44M,2020A&A...634A..53F,2022A&A...667A.101F,2023A&A...676A..33H} for more details).} We utilized Single Stellar Population (SSP) spectra from the eMILES library \citep{2016MNRAS.463.3409V}, covering metallicities from solar to -2.27 dex and ages from 70 Myr to 14.0 Gyr, with a Kroupa initial mass function (IMF, \citealt{2001MNRAS.322..231K}). The SSP library spectra were convolved with the line-spread function following \citet{2017A&A...608A...5G} and detailed in \citet{2019A&A...625A..76E}. 
A variance spectrum derived from the masked data cube was incorporated into pPXF to enhance the fitting process. For the kinematic fit, we used 8 degrees of freedom for the additive polynomial and 12 degrees for the multiplicative polynomial \citep{2019A&A...625A..76E}. In the age and metallicity fits, we constrained the velocity, omitted additive polynomials, and maintained the 12th degree in the multiplicative polynomial \citep{2019A&A...625A..77F}. We used pPXF weights to compute mean metallicities, mean ages, and stellar mass-to-light ratios from the SSP models for each galaxy. These stellar population properties are mass-weighted. To improve the fits, we masked residual sky lines not removed by ZAP. The spectra and best-fit models, together with some of the major absorption line regions (H$\beta$, Mgb, Fe, H$\alpha$, and CaT) are presented in Fig. \ref{fig:spectra}, and the derived properties are compiled in Table \ref{tab:sample}.

Uncertainties on the best-fit parameters were estimated using a Monte Carlo method, where residuals were reshuffled in a bootstrap approach, with uncertainties given as the 1$\sigma$ standard deviation of the posterior distribution. The signal-to-noise (S/N) ratio per pixel was calculated in a region between 660 and 680 nm, devoid of strong absorption or emission lines, as the mean fraction between flux and the square root of the variance, multiplied by the $\chi^2$ value estimated by pPXF. Since the main objective was to measure line-of-sight velocities, the targeted S/N ratio was 5, and caution is advised regarding age and metallicity estimates. {Based on tests of the recovery of SSP input parameters as a function of the S/N ratio with pPXF, \citealt{2019A&A...628A..92F} suggest that a S/N ratio of >10 is needed to determine metallicity within 0.2 dex with MUSE, and such ratios are achieved only for one target (i.e. KK\,218). For ages, an even larger S/N ratio of >15 is needed for a constraint within 1 Gyr, below it scatters with almost $\pm5$\,Gyr. Again, only KK\,218 achieves this S/N ratio (S/N $=$ 22). The other dwarfs have have age estimates between 6 to 12 Gyr and can therefore be considered of intermediate (2-8 Gyr) to old (> 8 Gyr) age, but we omit the values in Table\,\ref{tab:sample} and only show those which have a secured estimate.}

To find any compact objects associated with the dwarfs, namely globular clusters or nuclear star clusters, we  used the SEP segmentation map of all MUSE stacks as an input of x and y positions. For each segmented object, we used a small circular aperture and extracted the spectrum within, weighting the spectra of each spaxel with a PSF (approximated with a Gaussian) to boost the signal. We visually inspect each spectrum and its derived parameters. For KK\,208, we detect a globular cluster, which is also presented in Fig. \ref{fig:spectra}, and  in Table \ref{tab:sample}. No other compact stellar object associated with the dwarfs were found.

To estimate the structural properties of the globular cluster around KK\,208, we performed aperture photometry on Legacy Survey data \citep{2019AJ....157..168D} in $g$ and $r$ bands. The effective radius is estimated with modeling the object using a S\'ersic profile with Galfit \citep{2002AJ....124..266P}. We used two different stars in the MUSE field of view as PSF model. 

The known dwarf galaxy system surrounding M\,83 is listed in Table\,\ref{tab:group}.

\begin{table*}[ht]
\caption{The observed dwarf galaxies and globular cluster with MUSE.}             
\centering                          
\begin{tabular}{l c c l l c r c c l}        
\hline\hline                 
 Name &        RA &      Dec &         vel &        [M/H] &        age &         M/L & S/N & $M_V$ \\    
 & deg & deg & km/s & dex & Gyr & M$_\odot$ /L$_\odot$ &  & mag\\    
  (1) &        (2) &      (3) &  (4) &         (5) &        (6)&        (7) &        (8) & (9)   \\    
\hline      \\[-2mm]                  
dw1335-29 & 13:35:46.7 & $-$29:42:28 & 706.0$\pm$17.8 & $-$1.49$\pm$0.13 & --- & 1.91$\pm$0.11 &  6.7 & $-$10.3 \\
KK208 & 13:36:35.5 & $-$29:34:15 &  439.9$\pm$15.6 & $-$1.53$\pm$0.23 &  --- & 1.36$\pm$0.33 &  7.0 & $-$15.9 \\
dw1340-30 & 13:40:19.2 & $-$30:21:31 & 429.3$\pm$21.6 & $-$1.71$\pm$0.19 & --- & 1.23$\pm$0.27 &  9.4 & $-$10.8 \\
dw1341-29 & 13:41:20.2 & $-$29:34:03 & 483.4$\pm$20.6 & $-$2.21$\pm$0.13 &  --- & 1.16$\pm$0.36 &  8.5 & $-$8.8 \\
KK218 & 13:46:39.5 &  -29:58:45 &  594.9$\pm$9.0 & $-$1.68$\pm$0.08 &  8.3$\pm$1.4 & 1.68$\pm$0.18 & 22.3  & $-$12.1\\
    \\
KK208 GC1 & 13:36:25.90  & $-$29:35:29.2 &  432.1$\pm$3.3 & $-$1.48$\pm$0.04 &  13.7$\pm$1.6 & 2.19$\pm$0.20 & 33.3 & $-$5.5 \\
    
\hline
\end{tabular}
\tablefoot{ (7) the M/L ratio is the stellar mass-to-light ratio estimated by pPXF fitting of the spectrum. (8) the signal-to-noise is measured on the spectrum in a region between 660 and 680\,nm.}
\label{tab:sample}
\end{table*}

\begin{table}[ht]
\caption{The currently known dwarf galaxy system of M\,83.}             
\renewcommand{\arraystretch}{1.2}
\setlength{\tabcolsep}{2pt}
\centering                          
\begin{tabular}{cclrlc}        
\hline\hline                 
  Name &         RA &       Dec &  L$_V$ &        v &                 D \\
  &        J2000.0 &       J2000.0 &  10$^6$ L$_\odot$ &        km/s &                 Mpc \\
\hline      \\[-2mm]                  
         KK195 & 13:21:08.2 & $-$31:31:47 &             5.9 &  572$\pm$9 & $5.6^{+0.3}_{-0.2}$ \\
         KK200 & 13:24:36.0 & $-$30:58:20 &            21.5 &  494$\pm$9 & $4.8^{+0.1}_{-0.2}$ \\
        IC4247 & 13:26:44.4 & $-$30:21:45 &            85.5 &  420$\pm$9 & $5.2^{+0.1}_{-0.1}$ \\
     dw1335-29 & 13:35:46.7 & $-$29:42:28 &             1.1 & 706$\pm$17 & $5.0^{+0.3}_{-0.3}$ \\
       UGCA365 & 13:36:30.8 & $-$29:14:11 &            37.3 &  577$\pm$9 & $5.4^{+0.1}_{-0.1}$ \\
         KK208 & 13:36:35.5 & $-$29:34:15 &           195.9 & 439$\pm$15 & $5.0^{+0.2}_{-0.2}$ \\
      HID J1337-33 & 13:37:00.6 & $-$33:21:47 &             2.8 &  591$\pm$9 & $4.5^{+0.4}_{-0.2}$ \\
    ESO444-084 & 13 37 20.2 & $-$28:02:46 &            37.3 &  587$\pm$9 & $4.6^{+0.4}_{-0.4}$ \\
     dw1340-30 & 13:40:19.2 & $-$30:21:31 &             1.8 & 429$\pm$21 & $5.1^{+0.0}_{-0.1}$ \\
        IC4316 & 13:40:18.1 &  $-$28:53:40 &           102.8 &  576$\pm$9 & $4.3^{+0.1}_{-0.0}$ \\
     dw1341-29 & 13:41:20.2 & $-$29:34:03 &             0.3 & 483$\pm$20 & $3.9^{+1.4}_{-0.9}$ \\
         NGC5264 & 13:40:37.0 & $-$29:54:50 &           409.3 &  478$\pm$9 & $4.8^{+0.1}_{-0.2}$ \\
         KK218 & 13:46:39.5 & $-$29:58:45 &             5.9 &  594$\pm$9 & $4.9^{+0.2}_{-0.1}$ \\
\hline
\end{tabular}
\tablefoot{The distances are compiled in the LV catalog (D$<$10\,Mpc, \citealt{2013AJ....145..101K}) and are from the HST programs of \cite{2002A&A...385...21K,2007AJ....133..504K}, \cite{2003ApJ...596L..47P}, and \cite{2007MNRAS.374..107G} and the VLT program of \cite{MuellerTRGB2018}. Velocities are from \cite{1999ApJ...524..612B}, \cite{2004AJ....128...16K}, \cite{2008MNRAS.386.1667B}, \citep{2008AstL...34..832K}, and the analysis provided here. The luminosities come from \citet{2015A&A...583A..79M} and \citet{2024A&A...684L...6M}.}
\label{tab:group}
\end{table}

\section{The luminosity-metallicity relation}
\label{sec:lum_metal}
The stellar bodies of dwarf galaxies follow a relation in mass/luminosity and metallicity: the fainter, and thus, less massive a galaxy becomes, the more metal poor it is. This has been found {first for the dwarf irregular galaxies in the Local Group \citep{1989ApJ...347..875S} and later extended to nearby dwarf irregulars \citep{2006ApJ...647..970L}. \citet{2006ApJ...647..970L}  showed that the mass-metallicity relation established for large galaxies \citep{2004ApJ...613..898T} extends to the dwarf galaxy regime.   Using high quality data for Local Group dwarf spheroidal galaxies  \citet{2013ApJ...779..102K} showed that the mass-metallicity relation holds all the way from large galaxies to dwarfs also considering metallicities based on absorption line measurements (e.g. \citealt{2005MNRAS.362...41G}).} 

With modern facilities it is possible to estimate metallicities for dwarf galaxies outside our own Local Group. Most recently, \citet{2023A&A...676A..33H} observed 56 dwarf galaxies from the MATLAS survey \citep{2020MNRAS.491.1901H,2021MNRAS.506.5494P} with MUSE and found an offset with respect to the Local Group dwarfs luminosity-metallicity relation towards more metal-poor stellar populations. {Additional observations of 9 dwarf galaxies in the MATLAS survey with MUSE strengthened  this trend \citep{2025A&A...693A..44M}. However, it is not clear whether this has a physical or systematic cause due to the way the metallicities are measured (fitting CaT absorption  lines on individual stars or the multiple lines of the integrated stellar body of the galaxy).} 

\begin{figure}[ht]
    \centering
    \includegraphics[width=\linewidth]{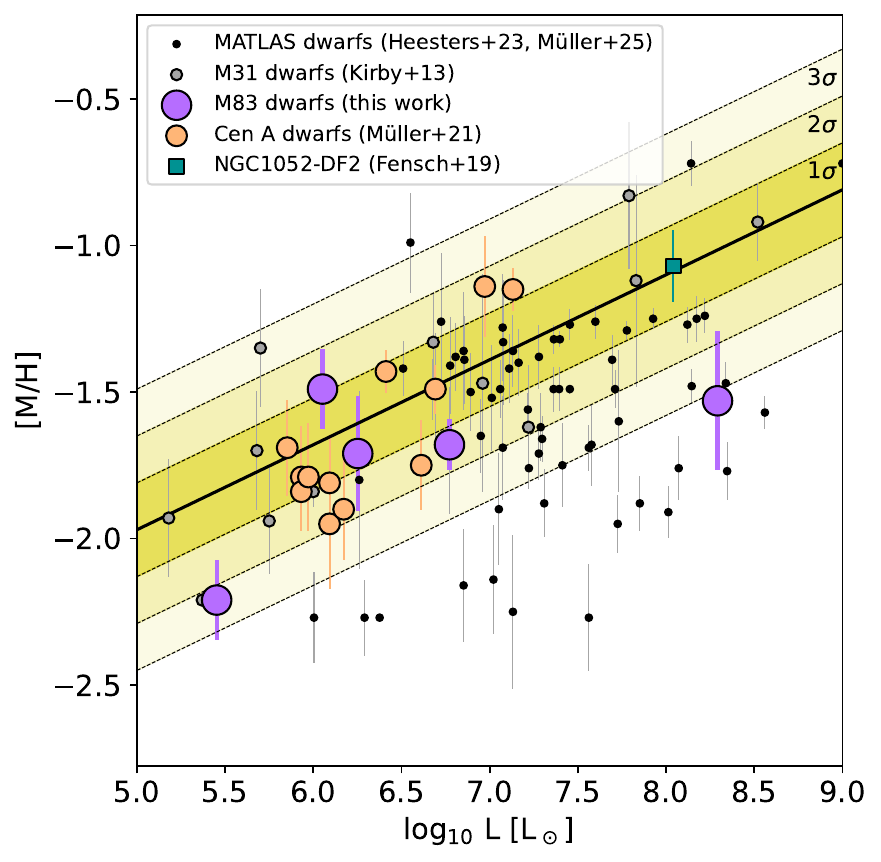}
    \caption{The luminosity-metallicity relation  for a reference sample of dwarf galaxies from the MATLAS survey \citep{2023A&A...676A..33H,2025A&A...693A..44M}, the ultra-diffuse galaxy NGC\,1052-DF2 \citep{2019A&A...625A..77F}, the Cen\,A  dwarfs \citep{Muller2021a}, and the here observed M\,83 dwarf galaxies. {The line and intervals indicate the fit by \citet{2013ApJ...779..102K} for the Local Group dwarfs and the 1, 2, and 3$\sigma$ intervals.}}
    \label{fig:mass_metal}
\end{figure}

We show the luminosity-metallicity relation of the M\,83 dwarfs  in Fig.\,\ref{fig:mass_metal} and compare it to the MATLAS dwarfs observed with MUSE, the M\,31 dwarfs, as well as to observations of Cen\,A dwarfs with MUSE \citep{Muller2021a}. We find that the M\,83 (and Cen\,A) dwarfs follow the luminosity-metallicity relation as expected from the Local Group dwarfs \citep{2013ApJ...779..102K}  within $\sim$3$\sigma$ (incl. the uncertainty). However, we note that in the mass range where the dwarf galaxies in the MATLAS survey deviate most from the relation (between 10$^{7.0}$ to 10$^{8.5}$\,L$\odot$) the observed M\,83 dwarf's  value is at the lower 3$\sigma$ limit.  However, this is KK\,208, which has a low S/N ratio and therefore large uncertainties. {Compared to the MATLAS dwarfs, the Centaurus dwarfs (which includes the Cen\,A dwarfs and the M\,83 dwarfs), seem to generally follow the luminosity-metallicity relation. This is striking, because the code used for the MATLAS and Centaurus dwarfs is essentially the same, indicating that the deviation may not come from systematics alone but rather is physical in nature.}

\section{The globular cluster of KK\,208}
\label{sec:glob}

\begin{figure}[ht]
    \centering
    \includegraphics[width=0.3\linewidth]{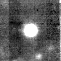}
        \includegraphics[width=0.3\linewidth]{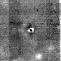}
            \includegraphics[width=0.3\linewidth]{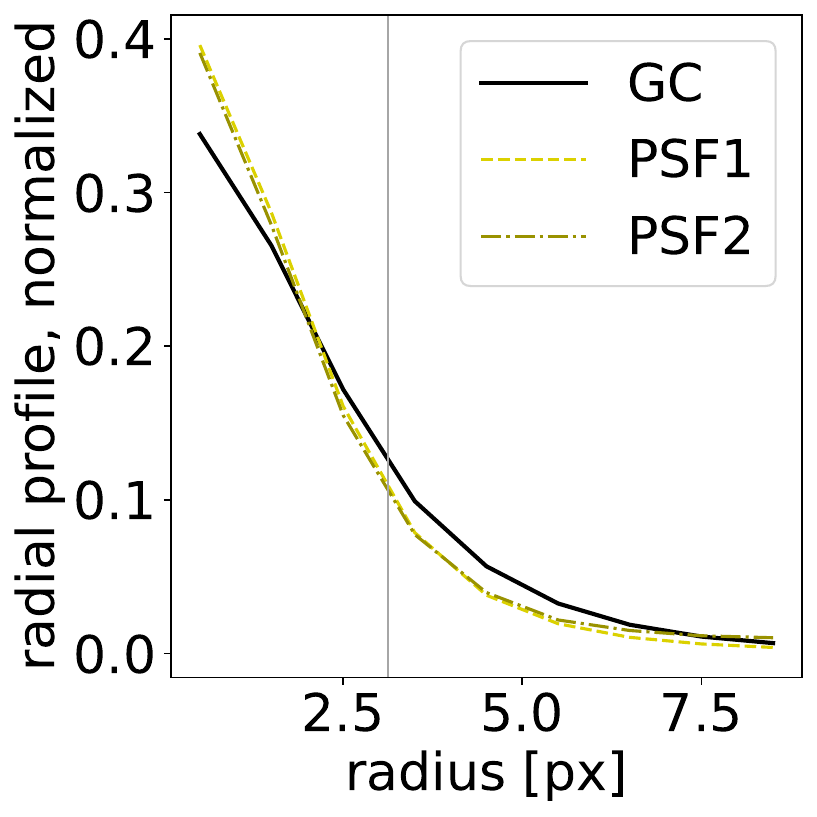}

    \caption{The globular cluster associated with KK\,208. Left: the MUSE stacked image. Middle: the S\'ersic model subtracted image. Right: the normalized radial profile of the globular cluster (black line) associated with KK\,208 and two PSF stars (orange dashed lines). The gray vertical line indicates the half light radius of the globular cluster estimated by Galfit.}
    \label{fig:gc}
\end{figure}

The velocity of the stellar body of KK\,208 is $ 439.9\pm15.6$ km/s and that of the globular cluster found in the MUSE pointing is  $ 432.08\pm3.3$ km/s. These values are consistent with each other. The globular cluster is therefore likely associated with the dwarf galaxy.  It seems unlikely that the globular cluster could be a contaminant star, according to the Besançon model of Galactic foreground stars \citep{2003A&A...409..523R}. The expected velocity distribution for the Milky Way stars in the direction of M\,83 is well approximated by the sum of two Gaussian profiles, with means $V_{Bes,1} = -3$\,km/s, $V_{Bes,2} = 76$\,km/s, and standard deviations $\sigma_{Bes,1} = 38$\,km/s, $\sigma_{Bes,2} = 102$\,km/s. 
Therefore, the objects velocity is on the tail end of the velocity distribution, more than 3 sigma above the mean. The line-of-sight velocity of M\,83 is $v_{M83}=519\pm18$\,km/s, which is further away from the velocity of the globular cluster. While still reasonable to assume that it could be a GC in the halo of M\,83, the closeness in phase-space to KK\,208 makes this unlikely and we conclude that it is a GC associated with KK\,208.

In Fig.\,\ref{fig:gc} we show the white image of the GC and the S\'ersic subtracted residual image. We note that there is a residual in the center, which is typical for slightly extended objects. In Fig.\,\ref{fig:gc} we also show the normalized radial profile of the GC, as well as a radial profile of two PSF stars. It is evident that the globular cluster is marginally resolved with more extended wings compared to the PSF. This is expected. GCs at the distance of Cen\,A/M\,83 are partly resolved showing broader and less peaked light distributions when compared to stellar PSF (see e.g. Fig\,1 in \citealt{2001A&A...369..812R}).  

We measure the size of the GC using 
\begin{equation}
    FWHM_{intr}= \sqrt{FWHM_{GC}^2-FWHM_{PSF}^2},
\end{equation}
and
\begin{equation}
    \sigma_{GC}=FWHM_{intr}/2.355,
\end{equation}
where $FWHM_{GC}=0.40$\,arcsec and  $FWHM_{PSF}=0.33$\,arcsec correspond to  the measured full width half max of the GC and PSF, respectively. At the distance of KK\,208 this translates into 5.6\,pc. Calculating $\sigma_{GC}$ from this we estimate an effective radius of 2.4\,pc for the GC. At a distance of 5.01\,Mpc for KK\,208 \citep{2021AJ....162...80A} and correcting for galactic extinction, we find an absolute magnitude of $-$5.5 mag in the $V$ band for the globular cluster. The size and brightness of the GC  are typical for such objects \citep[e.g., ][]{2009MNRAS.392..879G}.

\begin{figure}[ht]
    \centering
    \includegraphics[width=\linewidth]{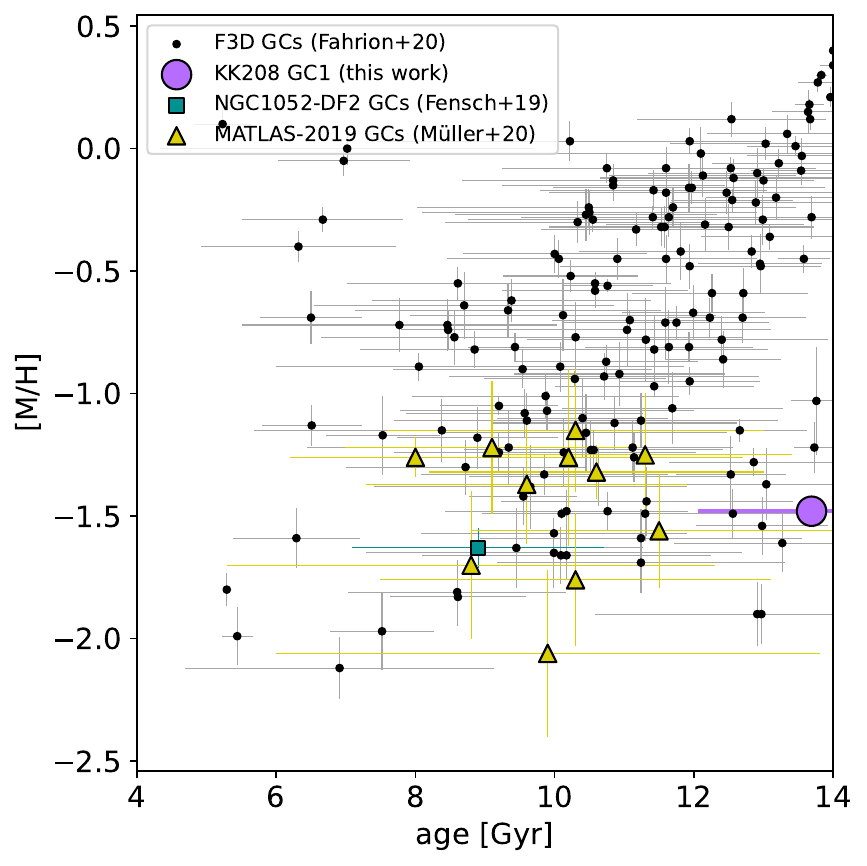}
    \caption{The age-metallicity relation for a reference sample of globular clusters from the F3D survey \citep{2020A&A...637A..26F}, the ultra-diffuse galaxy MATLAS-2019 \citep{2020A&A...640A.106M}, the stacked clusters of the ultra-diffuse galaxy NGC\,1052-DF2 \citep{2019A&A...625A..77F}, and the here observed globular cluster of KK\,208.}
    \label{fig:gc_age_metal}
\end{figure}

{With a metallicity estimate of $[M/H]=-1.48 \pm 0.04$\,dex and an age estimate of $13.7 \pm 1.6$\,Gyr the globular cluster of KK\,208 is consistent with measurements of other globular clusters observed with MUSE (see Fig\,\ref{fig:gc_age_metal}) and has a stellar mass of $3.0\times10^4$\,M$_\odot$. It is old and metal poor and has a similar metallicity as its host KK\,208 ($[M/H]=-1.53 \pm 0.23$\,dex).}

\section{The phase-space distribution of the satellites}
\label{kinematics}
Now that we have a complete set of dwarf galaxies around M\,83 with both confirmed membership from distances and velocities, we can study the phase-space distribution of the satellite system. 
\subsection{The dwarf galaxy population}

\begin{figure}[ht]
    \centering
    \includegraphics[width=\linewidth]{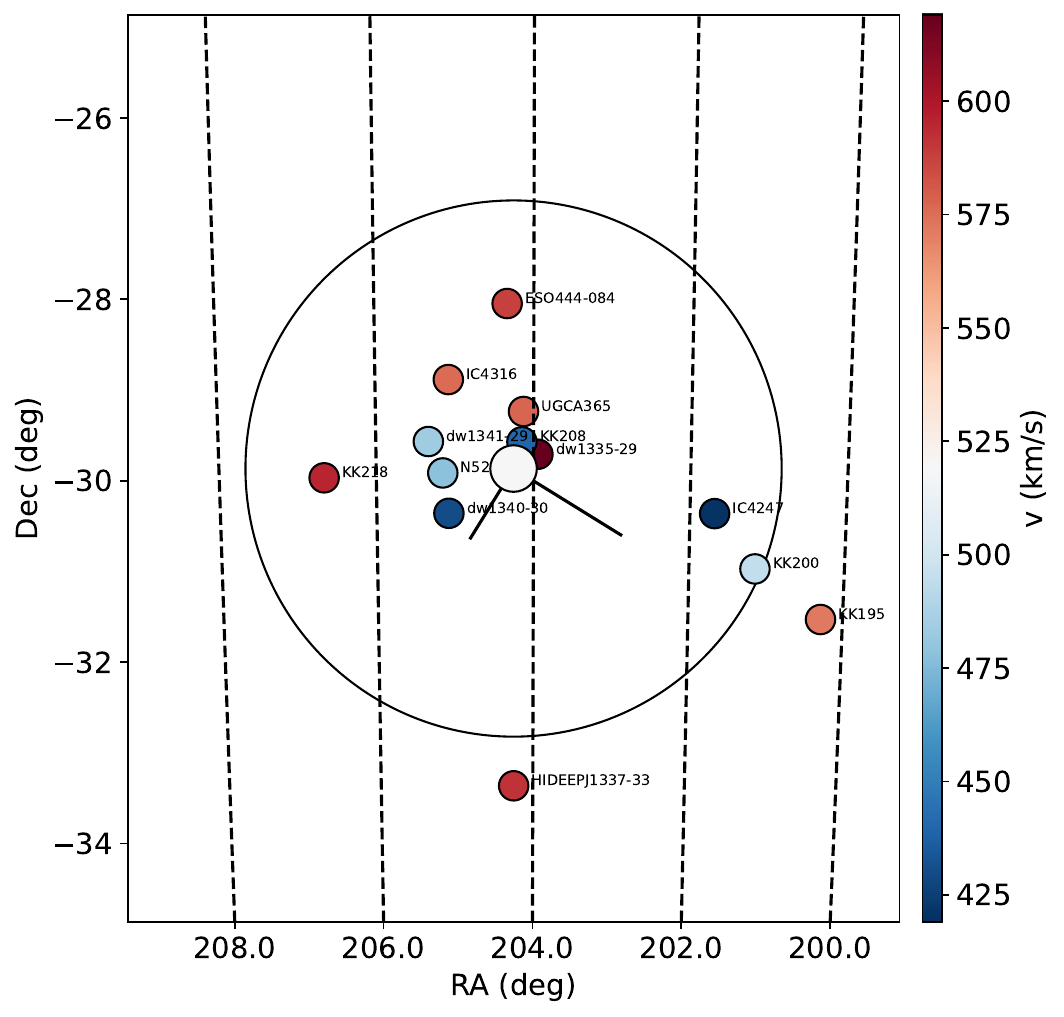}
    \caption{The field around M\,83 (large white dot) and its surrounding dwarfs (colored dots) in equatorial coordinates (J2000.0). The colors represent the line-of-sight velocity of the galaxies, as indicated with the color bar. {The two black lines correspond to minor-and-major axis of the satellite system (see text), and the circle to the virial radius.}}
    \label{fig:field}
\end{figure}

\begin{figure}[ht]
    \centering
    \includegraphics[width=\linewidth]{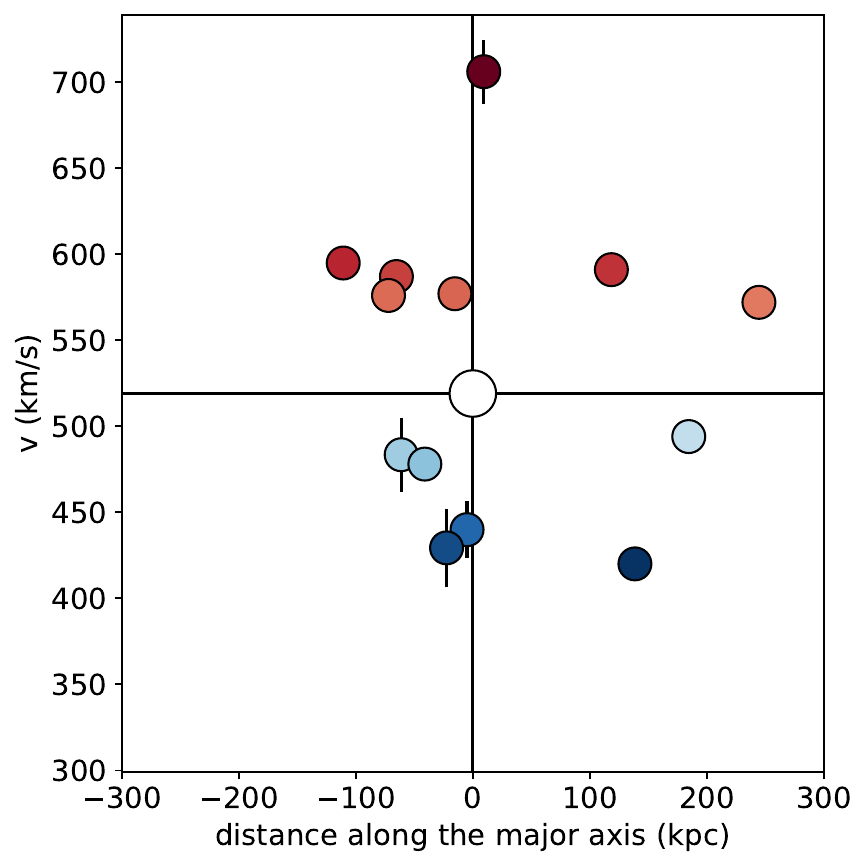}
    \caption{The position velocity diagram of the satellite system (small colored dots) of M\,83 (large white dot). The x-axis corresponds to the projected distance along the major axis of the satellite system, the y-axis to the line-of-sight velocities. The color coding corresponds to the velocity and is the same as for Fig.\,\ref{fig:field}.}
    \label{fig:pv_diagram}
\end{figure}

{\citet{MuellerTRGB2018} studied the spatial distribution of the satellites around M\,83 and noted that a 3D analysis is unfeasible compared to the studies in the Local Group and Cen\,A. This is due to the fact that the average distance uncertainties are of the order of $\approx$300\,kpc, which is larger than the virial radius of the host. Therefore we will solely work in projected distances.}
First, we estimate the geometry. For that we apply a principle component analysis, which returns the eigenvectors corresponding to the major and minor axis of the projected satellite distribution. These vectors are indicated in Fig.\,\ref{fig:field}. The minor-to-major axis ratio is 0.65, with a semi-major axis length of 111\,kpc. As seen in Fig.\,\ref{fig:field}, there are two dwarf galaxies outside of the virial radius (see Sec.\,\ref{sec:mass} for the calculation of the radius): KK195 and HIDEEPJ1337-33. While the former is within the extension of the major axis, the latter is almost perpendicular to it and rather isolated. Removing the two, and repeating the analysis {(effectively studying the dwarfs within the virial radius)}, we get a semi-major axis of 92\,kpc and a ratio of 0.44. In both cases, the flattening is not strong.

We now investigate the motion of the satellites within the system. For that, we project their distances from the minor axis (derived from the full sample)  along the major axis, centered on M\,83. As a convention, we assign negative distances to the left side of the minor axis (i.e. in direction of larger RA). In Fig.\,\ref{fig:pv_diagram} we plot the line of sight velocity as a function of the projected distance and split the figure in four quadrants. A fully rotationally supported system will occupy two opposing quadrants, a fully pressure supported and isotropically distributed system all four equally. For the M\,83 we find no signal for a rotationally supported system, with 7/13 and 6/13 satellites being on opposing quadrants, respectively. To summarize, we do not find a significant flattening, nor signals of co-rotation for the M\,83 satellite system aligned with this flattening.

The dwarf galaxy dw1335-29 has a higher velocity than the rest of the system. This is evident in Fig.\,\ref{fig:pv_diagram}. 
The 1$\sigma$ standard deviation of the the velocity of the full satellite system is 81\,km/s. With a relative velocity to M\,83 of 187\,km/s, dw1335-29 is at the 2.3$\sigma$ tail of the distribution. It is the closest system to M\,83, with a projected separation of only 27\,kpc and may therefore be on or close to its pericenter. Notably, the tidally disrupted dwarf KK\,208 is close by in projection to dw1335-29 (see Fig.\,10 in \citealt{MuellerTRGB2018}), however, the two have a velocity difference of 266\,km/s, making an association unlikely.

\subsection{A satellite-of-satellite system}

Two dwarf galaxies are close in phase-space. They are NGC5264 and dw1341-29. Their projected separation is only 32\,kpc and they share a common velocity, $478\pm9$\,km/s and $483\pm20$\,km/s, respectively. NGC\,5264 has an absolute $V$ band magnitude of $-$16.7 ($4.1\times10^8$\,L$_\odot$) and is the brightest satellite of M\,83, while dw1341-29 is so far the faintest discovered satellite, with $-$8.8 mag ($2.8\times10^5$\,L$_\odot$). They have a luminosity ratio of 1445. This pair may be a satellite of a satellite system. Let us quantify this.

\citet{2010ApJ...711..361G} provided a criteria for a pair of satellites to be bound when  
 \begin{equation}
      b \equiv \frac{2GM_\mathrm{pair}}{\Delta r\Delta v^2} > 1,
 \end{equation}

 where  $\Delta r$ is the physical separation between the objects, $\Delta v$ is their total velocity difference,  $M_\mathrm{pair}$ is the total mass of the pair, and  $G$ is the gravitational constant. Using this formula, we can solve for the minimal mass $M_\mathrm{pair}$ to be bound by assuming $b\approx1$.  We find that  $M_\mathrm{pair} > 9.3\times10^7$\,M$\odot$ to be considered bound. Including the uncertainties in distance (taking  half the projected distance as uncertainty in depth) and velocity, we find a conservative criteria of $M_\mathrm{pair} > 2.6\times10^9$\,M$\odot$. For the former, the stellar mass of NGC5264 ($8\times10^8$\,M$_\odot$, using a stellar M/L ratio of two to transform the luminosity into a stellar mass) alone is large enough to satisfy this criteria. For the latter, the pair would need only three times the observed stellar mass in form of gas or dark matter to satisfy the criteria. This is realistic (see e.g. Fig. 11 in \citealt{2012AJ....144....4M}), with dwarf galaxies at this luminosity hosting such dark matter halos. We therefore conclude that dw1341-29 is a bound satellite of NGC5264 and that the two represents a satellite-of-satellite system.

\section{Mass estimations of the M\,83 group}
\label{sec:mass}
In the following we compare different methods to derive the mass of the M\,83 halo. Since we did not detect any sign of coherent rotation of the satellite system, we do not include any correction for possible rotation when using the satellites as tracers.

\subsection{Virial theorem}

Previously, the mass was estimated by \citet{2002A&A...385...21K,2007AJ....133..504K} based on the virial theorem. Using their equation 
 \begin{equation} M_{vir}= 3\pi \frac{N}{(N-1)G}\sigma_{los}^2 \langle R_{ij} \rangle,
 \end{equation}
where $N$ is the number of tracers, $G$ the gravitational constant, $\sigma_{los}$ the measured velocity dispersion along the line-of-sight, and $\langle R_{ij} \rangle$ the mean  pairwise projected separation between the dwarf galaxies. For the M\,83 group, we find higher values for both $\sigma_{los}$ and $\langle R_{ij} \rangle$ compared to \citet{2007AJ....133..504K}. We estimate  $\sigma_{los}=80.6 \pm 15.4 $\,km/s and $R_H=163$\,kpc (compared to 61\,km/s and 89\,kpc, respectively). This leads to a higher mass estimation for M\,83, namely $M_{vir}=2.5\pm0.7\times10^{12}$\,M$_\odot$ (compared to $0.8\times10^{12}$\,M$_\odot$, \citealt{2007AJ....133..504K}). The errors are estimated by Monte Carlo simulations: First a new set of velocities is created for each iteration by drawing the velocity from  a Gaussian distribution with the mean set as the observed velocity and the standard deviation corresponding to the velocity uncertainties. These new velocities are randomly assigned to the dwarfs. Then we sample from the dwarf galaxy catalog by drawing between 8 and 13 dwarfs for each iteration. The 1$\sigma$ bound in log space of the $ M_{vir}$ posterior give the uncertainties. 

\subsection{Density contrast}

Is the updated virial mass of M\,83 realistic? In a $\Lambda$CDM cosmological context, the masses of galaxy groups can be estimated for a density contrast $\Delta$ with respect to the critical density of the Universe $\rho_{\rm c}$. 
The radius $R_\Delta$ defines the volume where the density is equal to $\Delta \rho_{\rm c}$.
If we assume that the measured flat part of the rotation curve of M\,83 -- $v_{\rm flat}=190$\,km/s \citep{2021A&C....3400448D} -- corresponds to the circular velocity $v_{\rm circ}$ at radius $R_\Delta$, the total encompassed mass (baryons and dark matter) is given by \citep[see, e.g.][]{2012AJ....143...40M}:
 \begin{equation}
M_{\Delta}=(\Delta/2)^{-1/2}(GH_0)^{-1}v_{\rm circ}^3,
\end{equation}
where $H_0=75.1\pm3.8$\,km/s/Mpc \citep{2020AJ....160...71S} is the Hubble constant\footnote{{The value of the Hubble constant is still debated, but ranges between 67 and 75\ \citep{2019ApJ...882...34F,2020A&A...641A...6P,2020ApJ...902..145K,2021A&A...647A..72K,2021ApJ...908L...6R,2021ApJ...911...12J}. We adopt a Hubble constant derived from the baryonic Tully-Fisher relation, which favors higher values.}}, and $R_\Delta$ is given by
\begin{equation}
R_{\Delta}=\frac{G M_{\Delta}}{v_{\rm circ}^2}.
\end{equation}
Adopting the commonly used value of $\Delta=200$, we derive for the M\,83 group a mass of $M_{200}= 2.1\pm0.3\times10^{12}$\,M$_\odot$ within $R_{200}=253\pm0.3$\,kpc. The uncertainties are derived by assuming an uncertainty of 10\,km/s for the rotation curve measurement.
This mass estimate is consistent with the mass estimate following the virial theorem we derived before ($M_{vir}=2.5\pm0.7\times10^{12}$\,M$_\odot$) and is higher than the values suggested by \citet{2007AJ....133..504K}.

\subsection{Abundance matching}
Using abundance matching, we can directly estimate the halo mass from the stellar mass of M\,83. Using Table 3 of \citet{2010ApJ...717..379B} as a look-up table (and interpolating between their data points), we derive a halo mass for M\,83 of $M_{AM} = 3.0\pm0.5\times10^{12}$\,M$_\odot$. The error is derived from assuming a 10 percent uncertainty on the stellar mass of M\,83. This value is higher than the previous two estimates, but still within 1 to 2\,$\sigma$ and thus compatible.  

Using the abundance matching relation between the effective radius and $R_{200}$ \citep{2013ApJ...764L..31K} with an effective radius of M\,83 of 3.5\,kpc \citep{2021ApJS..255...19L}, we derive a virial radius of 233\,kpc, which is similar to the value estimated from the density contrast (253\,kpc). Now, if we estimate the halo mass from the virial radius through the relation

\begin{equation}
     \frac{\Delta H_0^2}{8\pi G} = \frac{M_{200}}{4\pi R_{200}^3},
\end{equation}
we can get another estimate for the mass through abundance matching. With $H_0$ and $\Delta$ as before, we find that $M_{AM,r_{eff}} = 1.7\pm0.5\times10^{12}$\,M$_\odot$. This is lower than the \citet{2010ApJ...717..379B} abundance matching halo mass by a factor of two. For the estimation of the uncertainties we used the error on the Hubble constant (3.8 km/s/Mpc, \citealt{2020AJ....160...71S}) and a 10 percent uncertainty on the effective radius. 

\subsection{Integrated Jeans-based analysis}
{Another approach to measure the mass of the M\,83 group is through direct Jeans analysis. This has the advantage that we have control over the mass profile and the velocity anisotropy and can factor in these unknown parameters in the error estimation. For a low number of discrete data points, it is however advisable to rather use an integrated version of the Jeans equations, which avoids binning the already sparse data. \citet{2010MNRAS.406..264W}  presented mass estimators to compute the enclosed mass of galaxy halos from samples of discrete positional and kinematical data of tracers, such as dwarf satellites. Among these estimators, there is one for using only projected positions and line-of-sight velocities. \citet{2011MNRAS.413.1744A} expanded on this work. Following their section 4.2, the enclosed mass is given as:}

\begin{equation}
    M_{out} \approx \frac{1}{GN} \sum_i^N \omega(R_i)v_{i}^2,
\end{equation}
{where $\omega(R)$ is a weighting function and $v_i$ is the line-of-sight velocity of the $i$th tracer. The weighting function is calculated as:}

\begin{equation}
    \omega(R) = \frac{R^\alpha}{I_{\alpha\beta} r_{out}^{\alpha-1}}
\end{equation}
and 
\begin{equation}
    I_{\alpha\beta} = \frac{\pi^{1/2}\Gamma(\frac{\alpha}{2}+1)}{4\Gamma(\frac{\alpha+5}{2})}
    \frac{\alpha+3 - (\alpha+2)\beta}{\alpha+3-2\beta},
\end{equation}
{where $\Gamma(x)$ is the gamma function, $\alpha$ the power index of the scale-free potential (with  $\alpha=1$ representing a point mass and $\alpha=0$ a logarithmic, i.e. truncated singular isothermal sphere), and $\beta$ the anisotropy parameter for the spherical system.  In an isothermal
system where the radial and tangential dispersions are equal, $\beta=0$. For systems in which
the orbits are predominantly radial the anisotropy parameter is $\beta>0$ and for predominantly
tangential orbits it is $\beta<0$. Here, we neither know the values for $\alpha$ nor for $\beta$. Assuming $\alpha=0$ and the $\beta=0.3$ based on simulated halos \citep{2013MNRAS.429.3079M}, we derive a mass of  $M_{out} = 1.5\pm0.1\times10^{12}$\,M$_\odot$ within $r_{out}=325$\,kpc (the largest projected separation of a dwarf galaxy to M\,83). The uncertainties are based on sampling the velocities of the tracers using their uncertainties around the value as a normal distribution. This, however, does not include varying $\alpha$ or $\beta$ values. If we allow $\alpha$ to vary between $-1.5$ and 0.5, and $\beta$ within a Gaussian distribution centered on $\log(-\beta+1)$ with $\beta=0.3$ with a standard deviation of 0.3\,dex, we can construct a distribution of possible masses. Its standard deviation informs us about the uncertainty. We estimate $ M_{out,possible} = 2.2\pm0.5\times10^{12}$\,M$_\odot$, with the nominal value giving the mean of the distribution.  If we calculate the mean density within a sphere with $r_{out}=325$\,kpc, it is below the critical density we previously calculated using the density contrast by a factor of three. This means that the virial mass must be lower than the value we estimated here, or in other words, the estimations are upper limits.

The value estimated through a Jeans-based analysis is still higher than previous literature values, but smaller than the three other approaches we have used in the previous two subsections. Within the uncertainties, all values are mutually consistent though.}

\subsection{Modified Newtonian Dynamics (MOND)}
In the context of alternative gravity scenarios, such as MOND \citep{1983ApJ...270..365M}, the missing mass is rather explained by a modification of the laws of gravity than by a dark matter particle \citep[e.g., ][and references therein]{Famaey2012}. In MOND  Newton's law of gravity is changed for accelerations below the threshold a$_0\approx10^{-13}$\,km/s$^2$. In a MOND universe one could ask what would be the dark matter inferred from observations when applying Newton's law of gravity (without any MOND effect). In a MOND cosmology this would be called phantom dark matter. \citet{2021ApJ...923...68O} explored the predictions of the phantom dark matter mass inferred from the baryonic content of a galaxy in such a scenario. Their Eq.\,10 gives a form to calculate the phantom dark matter mass encompassed within a radius r:
\begin{equation}
    M_{PDM}(r)= M_b \times \left(
    {\nu  \left(
    {\frac{GM_b}{r^2a_0}}
    \right) 
    } - 1
    \right), \mathrm{with}\,\nu(x) = \left({\frac{1}{4} + \frac{1}{x}}\right)^{1/2} + \frac{1}{2} 
\end{equation}

Following the steps in \citet{2019A&A...623A..36M} to estimate the baryonic mass of spiral galaxies from the $K$ band magnitude we estimate $M_b=0.6\times10^{11}$\,M$_\odot$ for M\,83. This yields a phantom dark matter mass at $R_{200}$ of $M_{PDM}=1.9\pm0.3\times10^{12}$\,M$_\odot$, with the uncertainties estimated from a Monte Carlo run where the baryonic mass is assigned an uncertainty of 30 percent and $R_{200}$ are sampled. This value is consistent with the derived masses for the M\,83 halo in the previous four subsections.

\section{Discussion and conclusions}
\label{sec:disc}
With the new MUSE observations of five dwarf galaxies in the vicinity of M\,83, we complete the line-of-sight velocity information for all 13 known dwarfs around M\,83. Together with tip of the red giant branch distances for all but one dwarf (dw1341-29, which only has surface brigthness fluctuation distances), this yields a complete picture of the M\,83 group. Spectroscopic analysis of the five dwarf galaxies reveals that their stellar populations follow the universal luminosity-metallicity relation. The brightest dwarf galaxy {(in terms of integrated luminosity)} -- the tidally disrupted dwarf KK\,208 -- lies at the 3$\sigma$ border of the relation, but we note that the signal to noise ratio is low for accurate spectroscopy. It is still noteworthy that \citet{2023A&A...676A..33H} found a systematic offset from the relation, on which KK\,208 would fall. Better spectroscopy of KK\,208 is needed to test this though.

For KK\,208 we  discovered a globular cluster in the field of view of MUSE. Based on its velocity, it is associated with KK\,208. Its high signal to noise ratio allowed us to derive accurate age and metallicity properties. It is old and metal poor, as expected from observations of other globular clusters around metal-poor dwarf galaxies.

Satellite of a satellites systems have been studied in \citet{2023A&A...673A.160M} with respect to cosmological expectations. There it was found that observed satellites of satellites are unusually bright (i.e. have a low luminosity ratio between the dwarf galaxy and its brightest satellite) when compared to state-of-the-art $\Lambda$CDM simulations \citep{2018A&A...616A..96R}, especially in the brightness ranges between 10$^6$ to 10$^8$\,L$_\odot$. The potential satellite pair discovered here -- NGC\,5264 and dw1341-29 -- is at the bright end of their inspected sample, and is likely gravitationally bound based on their common velocity and small separation. The observed pair resembles the model dwarf h021 and its satellite (from \citealt{2018A&A...616A..96R}), where the main satellite has a luminosity of $2.3\times10^8$\,L$_\odot$ and the luminosity ratio   with its brightest satellite is 865. To compare, NGC\,5264 has a luminosity of $4.1\times10^8$\,L$_\odot$ and is 1445 times brighter than dw1341-29. These observed properties are well within the spread of the simulated models.
We therefore find that this satellite-of-satellite pair is in agreement with $\Lambda$CDM expectations based on the simulations of \citet{2018A&A...616A..96R}. More confirmed satellite of satellite systems need to be studied to test whether they follow the expectations from standard cosmology or not.

The M\,83 group forms together with the Cen\,A group the Centaurus aggregation (or Centaurus group). Cen\,A with its satellite system is one of the best studied systems outside of our  Local Group and was found to host a co-rotating plane-of-satellites. Compared to high-resolution $\Lambda$CDM simulations, such co-rotating systems are outliers and should be rare. It is therefore natural to ask whether the M\,83 group hosts a similar structure. For M\,83 it was put forward that there might be a flattened structure based on the three dimensional distribution of its satellites (see Fig.\,9 in  \citealt{MuellerTRGB2018}), but at the same time the 3D distance uncertainties are as large as such a putative structure. Also, no kinematic study was conducted. Here, we are using the on-sky projection and motion of the 13 dwarf galaxies around M\,83 to study whether there might be a co-rotating plane-of-satellites around M\,83. We do not find convincing evidence for that though. Rather, the motion is uncorrelated and while there is some flattening of the system, it is well consistent with expectations from cosmological simulations. For example, \citet{2024A&A...683A.250M} finds for analogs of the M\,81 group in the TNG50 simulation a typical flattening between 0.4 and 0.8 for its 19 satellites with velocity and distance information. The direct comparison to M\,81 is valid because their stellar luminosity is similar ($\log K=10.95$ for M\,81 and $\log K=10.86$ for M\,83, according to the updated Local Volume catalog, \citealt{2004AJ....127.2031K}) and the halo mass range for the selection in TNG50 was set to be between 0.5 to 2.0 $\times 10^{12}$\,M$_\odot$, which is the range of masses we here find for the M\,83 halo. 
With a measured flattening of 0.6 for the M\,83 satellite system it is within 1$\sigma$ from expectations (see Fig.\,6 in \citealt{2024A&A...683A.250M}). We conclude that current observations do not support a scenario where M\,83 hosts a co-rotating plane-of-satellites seen edge-on and that the system is well within  $\Lambda$CDM expectations. {The Cen\,A/M83 pair of galaxies is therefore different from the Milky Way/Andromeda galaxy pair when it comes to phase-space correlations. 

There are some caveats to consider concerning the non-detection of a coherently moving plane-of-satellites. We will only be able to detect a co-rotating plane if it is a) well-populated and b) observed under a low inclination. For example, only half the satellites of the Andromeda galaxy make up the co-rotating plane-of-satellites. The velocities of the dwarfs outside the plane are uncorrelated in phase-space. If we were to include all dwarf galaxies of the Andromeda galaxy in an analysis such as we performed, we would draw similar conclusions as for M\,83 (i.e. finding no co-rotating structure). Only by studying the subpopulation of the Andromeda satellite system the phase-space correlation becomes significant. So why are we not considering subpopulations around M\,83? This is due to sample size. While around the Andromeda galaxy there are more than 20 dwarf galaxies, here we have a sample of 13 dwarfs. The binomial chance of finding a correlated signal with half the population, this is, 6 dwarfs, would be 3 percent and would not allow us to draw any conclusions at a 3$\sigma$ level. So while the subpopulation of dwarfs around the Andromeda galaxy is large enough (15 dwarf galaxies are in the plane) to establish a phase-space correlation, the same is currently not given for M\,83. The other caveat is the inclination. Only when seeing a co-rotating plane-of-satellites almost edge-on, such as it is the case for the Andromeda galaxy or Cen\,A, are we able to detect them in the first place. A face-on plane or one with a high inclination will be difficult, albeit not impossible to detect. There is on-going efforts to be able to detect face-on planes-of-satellites from line-of-sight velocity measurements and it should be in principle be possible to distinguish a co-rotating plane-of-satellites seen edge-on from a pressure supported system by the absolute value of the velocity dispersion (Crosby et al., in preparation).

It is interesting to note that the planes-of-satellites around Cen\,A and the Andromeda galaxy are aligned with the nearby large-scale structure \citep{2019MNRAS.490.3786L}. In their work, the putative plane of M\,83 stood out as an outlier and no alignment with the shear tensors which define the directions of collapse were found. This is now unsurprising, as the existence of the putative M\,83 plane is not supported anymore by current data. Other outliers in their analysis were the Milky Way plane, for which we have excellent data which support the existence of a plane \citep{Pawlowski2018}, and M\,101 \citep{Muller2017}, which was further studied by \citet{2018AJ....156..105A} and appears to be a filament stretching out from the Local Sheet (see their Fig. 7) and extending several megaparsecs towards NGC\,6946 and beyond. It is therefore wrong to conclude that an outlier from the alignment with the shear tensors necessitates that the structure is not real.

With the velocities at hand for the whole satellite system, we can estimate the mass of the M\,83 group. Applying the virial theorem we estimate that the M\,83 group has a total mass of $2.5\pm0.7$ $\times$ 10$^{12}$\,M$_\odot$. This mass is consistent with the expected mass derived from the flat part of the rotation curve of M\,83  ($2.1\pm0.3\times10^{12}$\,M$_\odot$) and from abundance matching ($3.0\pm0.5\times10^{12}$\,M$_\odot$). The face value is larger than previous estimates but follows from a larger measured velocity dispersion. {Using instead an estimator based on Jeans modelling and reasonable assumptions for the velocity anisotropy ($\beta=0.3$) and an isothermal mass profile we estimate a mass of $1.5\pm0.1\times10^{12}$\,M$_\odot$. However, when sampling $\alpha$ and $\beta$ with realistic values, the masses are larger with  $2.2\pm0.5\times10^{12}$\,M$_\odot$.}
At a 2$\sigma$ level, all these measurements of the halo mass of M\,83 are in line with measurements from our own Milky Way group, ranging between 1.0 to 1.5 $\times$ 10$^{12}$\,M$_\odot$\citep{2014A&A...562A..91P,2019A&A...621A..56P,2022ApJ...925....1S,2024arXiv240506017K}. The halo mass  values for M\,83 estimated here are larger than previously assumed, which would have an impact on a recent study of the number of expected satellites around M\,83 (see next paragraph). The average halo mass from different tracers estimated here is $2.1\pm1.0\times10^{12}$\,M$_\odot$. This yields a dynamical mass-to-light ratio for the M\,83 group of 30, which indicates a dark matter dominated halo.
As a test for alternative gravity models, namely MOND, we calculated the expected phantom dark matter mass in such a scenario. With a phantom dark matter mass of   $1.9\pm0.3\times10^{12}$\,M$_\odot$, MOND correctly predicts the inferred total mass of M\,83 within the uncertainties. 

In \citet{2024A&A...684L...6M} the authors estimated that the count of 13 dwarf galaxies is too large in $\Lambda$CDM based on comparisons to cosmological simulations and predictions from the subhalo mass function. Similar findings have been reported for a statistical sample from the MATLAS survey \citep{2024arXiv240505303K} and our own Milky Way system \citep{2023arXiv231105439H}. 
For the comparison to cosmological simulations, \citet{2024A&A...684L...6M} used only the baryonic mass of M\,83 to select the halos within TNG50 of the Illustris suite (not the halo mass), therefore the updated total halo mass will not affect their results {and the tension remains}. 
For the theoretical predictions from the subhalo mass function, they however assumed  halo masses for M\,83 between 0.8 to 1.0  $\times$ 10$^{12}$\,M$_\odot$ and found a 5$\sigma$ to 3$\sigma$ discrepancy, respectively, to theoretical predictions. The larger the halo mass, the higher the number of expected satellites. A halo mass more than two times as large as used in \citet{2024A&A...684L...6M}  will move the tension below the 3$\sigma$ limit. 

It is critical to get an accurate halo mass estimation of M\,83 for such kind of studies. Recently, \citet{2023ApJ...947...34H}, and \citet{2024A&A...685A.132D} used different tracers such as dwarf galaxies, streams, and globular clusters to update the virial mass of the Cen\,A group from previous estimations using only the dwarf galaxies \citep{2002A&A...385...21K,2007AJ....133..504K,2022A&A...662A..57M}. Such an analysis can be extended to the M\,83 group. Especially the stream KK\,208 may constrain the halo mass of M\,83, as it was done for Cen\,A \citep{2022ApJ...941...19P}. Such an effort might push down the uncertainties and yield more accurate estimations of the halo mass of M\,83, which in turn will help constrain cosmological models. Furthermore, wide-area searches for globular clusters and spectroscopic radial velocity follow-up survey can provide additional constraints for a detailed cosmological mass modelling of M\,83.

\begin{acknowledgements} 
{We thank the referee for the constructive report, which helped to clarify and improve the manuscript.}
O.M. and N.H. are grateful to the Swiss National Science Foundation for financial support under the grant number  	PZ00P2\_202104. 
M.S.P. acknowledges funding of a Leibniz-Junior Research Group (project number J94/2020). This project has received funding from the European Union’s Horizon Europe research and innovation programme under the Marie Sk\l{}odowska-Curie grant agreement No 101103830. {SP was supported by a research grant (VIL53081) from VILLUM FONDEN.}

\end{acknowledgements}

\bibliographystyle{aa}
\bibliography{aa}

\end{document}